%% file: main.tex
\documentclass{article}
\usepackage{iclr2025_conference,times}

\input{preamble}

\title{
\centering
\Large Audio Flamingo Sound-CoT Technical Report: Improving Chain-of-Thought Reasoning in Sound Understanding
}
\author{
    \textbf{Zhifeng Kong, Arushi Goel, João Felipe Santos, Sreyan Ghosh, Rafael Valle,}\\
    \textbf{~Wei Ping, Bryan Catanzaro}\\
    ~~\\
    ~NVIDIA \\ \vspace{2mm}
    \texttt{zkong@nvidia.com}
}

\iclrfinalcopy

\begin{document}

\maketitle
\begin{abstract}
Chain-of-thought reasoning has demonstrated significant improvements in large language models and vision language models, yet its potential for audio language models remains largely unexplored. In this technical report, we take a preliminary step towards closing this gap. For better assessment of sound reasoning, we propose \texttt{AF-Reasoning-Eval}, a benchmark targeting common‑sense reasoning and the ability to discriminate among closely related choices. To prepare training corpus for sound reasoning abilities, we propose automatic pipelines that transform existing audio question–answering and classification data into explicit reasoning chains, yielding \texttt{AF-CoT-Train} with 1.24M samples. We study the effect of finetuning Audio Flamingo series on \texttt{AF-CoT-Train} and observe considerable improvements on several reasoning benchmarks, validating the effectiveness of chain-of-thought finetuning on advanced sound understanding. 
\end{abstract}

\section{Introduction}

In recent years, there have been significant advances in Audio Language Models (ALMs). These models can understand different types of audio -- including sound, speech, and music -- in terms of sound semantics, temporal orders, long-form structure, transcriptions, and so on \citep{deshmukh2023pengi,gong2023listen,tang2023salmonn,ghosh2025audio,chu2024qwen2,ghosh2024gama,xu2025qwen2_5_omni}. Similar to the Vision Language Model (VLM) area \citep{liu2023visual}, in ALMs users can input audio pieces and text prompts as instruction (e.g. a question, request to summarize, request to transcribe, and so on), and the ALM will output the answer in natural language. ALMs are usually built upon open-sourced Large Language Models (LLMs), where audio inputs are represented by an audio encoder and then fused into the LLM with cross attentions \citep{kong2024audio,ghosh2025audio} or self attentions \citep{deshmukh2023pengi,tang2023salmonn,xu2025qwen2_5_omni}. 

Most of these ALMs directly output the answer for an instruction. In contrast, studies in LLM have demonstrated that outputting intermediate thinking process, namely chain-of-thought (CoT) or reasoning chains, can significantly improve the accuracy across all tasks \citep{guo2025deepseek,bercovich2025llama}. CoT breaks a complex task into simpler and manageable tasks, and adaptively allocates different computation budgets based on the difficulty of the question, thus making the prediction more accurate and transparent. Furthermore, a series of works have introduced CoT reasoning to VLMs \citep{zhang2023multimodal,shao2024visual,xu2024llava,zhang2024improve}. Nevertheless, there are only very few studies on CoT in ALMs \citep{ma2025audio,xie2025audio,wijngaard2025audsemthinker}, leaving a huge gap in the understanding of several key questions
\begin{itemize}
    \item Does chain-of-thought reasoning improve the accuracy, robustness, and transparency of audio language models? If so, where do the improvements come from? 
    \item How do we curate training data to enable accurate chain-of-thoughts in audio language models? Should we use synthetic methods, manual annotations, or a hybrid approach with human in-the-loop? 
    \item Given the complication in the variety of audio-specific data and tasks, what is the optimal recipe to curate chain-of-thought training data and combine with existing training data, and what is the best method to train the reasoning abilities?
    \item How to decompose reasoning in the audio and text domains, and how to reliably measure the reasoning quality? 
\end{itemize}

In this technical report, we aim to conduct preliminary investigation to the above questions. Specifically, we study if directly finetuning ALMs on synthetic CoT data could improve their reasoning abilities and therefore accuracies. 
We restrict to sound understanding because this modality requires semantic and common-sense reasoning, and includes diverse audio and question types compared to speech and music. We focus on multiple-choice style question answering and classification for straight-forward evaluation. 

The first challenge is to curate high-quality audio CoT data for finetuning. Since human-annotated samples are expensive to obtain and hard to scale-up, we focus on synthetic CoT data. While there already exist some synthetic audio CoT datasets \citep{xie2025audio,wen2025sari,wijngaard2025audsemthinker}, they represent audio with caption or metadata and generate reasoning chains with text-only LLMs. These approaches potentially ignore the specific reasoning needs in the audio domain. For instance, these methods might not be able to generate complex reasoning chains that require understanding of relationships between audio events if these were not extracted by the audio captioning model. To address this challenge, we propose different CoT data generation pipelines with more involved LLM and ALM interactions to ensure the reasoning chain includes both audio-specific and text-specific reasoning. A central LLM breaks the complex question into smaller ones and queries the ALM at each step, ensuring there is audio-specific reasoning throughout our entire generated CoT data. We also apply several validation and filtering strategies to ensure the quality of generated reasoning chains. 
With these methods, we add about 1.24M reasoning chains to existing data, and name it \texttt{AF-CoT-Train}.

The second challenge is to conduct holistic evaluation of the model's reasoning abilities. In addition to existing benchmarks \citep{sakshi2024mmau,ma2025mmar}, we provide two additional reasoning benchmarks to foster the research in this area. The first evaluation set contains 150 human annotated question-answer samples with a focus on common sense reasoning. The second evaluation set contains 7227 challenging classification samples where the options are automatically generated to be similar to each other. These two subsets are collectively called \texttt{AF-Reasoning-Eval}. 

We finetune our prior works on ALMs, Audio Flamingo 2 \citep{ghosh2025audio} and Audio Flamingo 3 \citep{goel2025audio}, on \texttt{AF-CoT-Train} to study the effect of CoT finetuning. The resulting models are named Audio Flamingo 2 Sound-CoT and Audio Flamingo 3 Sound-CoT, respectively. We observe improvements on accuracies, and set new state-of-the-art on some of the benchmarks. Especially, Audio Flamingo 2 Sound-CoT, despite based on a small 3B LLM, outperforms open-sourced 7B baselines and closed-sourced Omni models on several benchmarks. We then conduct ablation studies on data and training recipes and found a set of good recipes. We also manually measure the quality of reasoning outputs and causality (i.e. whether the model follows reasoning chains when predicting) to decompose and analyze the success and failure patterns of our CoT finetuned models. Our project is released at: \url{https://github.com/NVIDIA/audio-flamingo/tree/soundCoT}.

\section{Related Works}

\subsection{CoT Reasoning in Large Language Models and Vision Language Models}

Chain-of-Thought (CoT) in an LLM is defined as a series of intermediate steps in the natural language form that decompose a complex question into manageable steps \citep{wei2022chain}. Different CoT methods -- including prompting, finetuning, and distillation -- help the LLM focus on each sub-problem at a step and allocates additional computation for harder problems, therefore remarkably improving the reasoning abilities and generation quality especially in terms of complex problem solving \citep{wei2022chain,zhang2022automatic,wang2022self,yao2023tree,guo2025deepseek,bercovich2025llama,liu2025acereason}.

Recently, several works have introduced CoT to Vision Language Models (VLMs) training and evaluation, including Multimodal-CoT \citep{zhang2023multimodal}, CURE \citep{chen2023measuring}, Visual-CoT \citep{shao2024visual}, LLaVA-CoT \citep{xu2024llava}, VCTP \citep{chen2024visual}, LLaVA-Reasoner, \citep{zhang2024improve}, M$^3$CoT \citep{chen2024m}, and Cosmos-Reason-1 \citep{azzolini2025cosmos}. Compared to CoT in LLMs, there are several unique challenges of CoT in VLMs: 
(1) CoT in VLMs includes reasoning in both the vision and the text domain, thus adding more complexity to training;
(2) while LLMs may have emergent CoT abilities already, VLMs usually need to be explicitly trained on CoT data;
(3) in addition to text-only failure cases in LLMs, there are more failure cases of CoT in VLMs with respect to task understanding, visual grounding, visual hallucination, and spatial understanding, leading to unique challenges in training and evaluation. 

The above works address these challenges through custom reasoning templates and well-designed CoT data generation and model training methods. For example, LLaVA-CoT \citep{xu2024llava} proposed a four-step template consisting \texttt{summary, caption, reasoning, conclusion} steps for explicit image tracing. In this paper, we apply this template to ALMs due to its direct applicability to the audio domain. Other works also propose multi-stage reasoning templates and use visual grounding information to foster reliable reasoning \citep{zhang2023multimodal,chen2024visual,shao2024visual}, or use RL to reduce hallucination \citep{zhang2024improve}.

\subsection{Audio Language Models}

Audio Language Models (ALMs) are a type of multimodal language models that take audio and text prompts as inputs and return text answers, similar to vision language models \citep{alayrac2022flamingo,liu2023visual}. The main tasks of audio language models include audio understanding (captioning, question answering, classification, etc.), speech understanding (recognition, translation, classification, etc.), and music understanding (information retrieval, tagging, structure analysis, etc.). The research community has been developing more general and capable ALMs since the first general-purpose ALM called Pengi \citep{deshmukh2023pengi}. The representational works along this line includes: LTU series \citep{gong2023listen, gong2023joint}, Salmonn \citep{tang2023salmonn}, Audio Flamingo series \citep{kong2024audio,ghosh2025audio,goel2025audio}, Qwen-Audio series \citep{chu2023qwen,chu2024qwen2}, GAMA \citep{ghosh2024gama}, Mu-LLaMA \citep{liu2024music}, MusiLingo \citep{deng2023musilingo}, Mellow \citep{deshmukh2025mellow}, Aero-1 Audio \citep{li2025aero}, and Kimi-Audio \citep{ding2025kimi}. There are also multimodal LLMs with audio understanding capabilities including Phi-4 Multimodal \citep{abouelenin2025phi}, Baichuan-Omni-1.5 \citep{li2025baichuanomni}, Qwen2.5-Omni \citep{xu2025qwen2_5_omni}, GPT-4o \citep{hurst2024gpt}, and Gemini \citep{team2023gemini,team2024gemini,comanici2025gemini}.

\subsection{CoT Reasoning in Audio Language Models}

The challenges in VLM CoT also apply to ALM CoT due to the similarity of these two domains. Additionally, it is more challenging to curate high-quality CoT training data for ALM due to the limited amount of data labeling tools and the scarce of data availability.
Recent studies show test-time compute scaling via CoT can improve model's accuracies \citep{ma2025audio, dang2025scaling}. However, the improvements are marginal on standard benchmarks such as MMAU \citep{sakshi2024mmau}, likely because the model has weak CoT reasoning abilities. Audio-Reasoner \citep{xie2025audio}, SARI \citep{wen2025sari}, and AudSemThinker \citep{wijngaard2025audsemthinker} generate audio reasoning datasets for SFT and RL, and show considerable improvements on MMAU. However, their audio reasoning data generation pipelines only represent audio with textual descriptions and over-rely on the reasoning abilities of a text-only LLM backbone. The lack of ALM and LLM interactions could lead to inaccurate reasoning chains. 

In this paper, we propose four different audio reasoning data generation pipelines, two for audio QA, and two for audio classification. Our pipelines leverage the interaction between text-only LLM and ALM especially in the reasoning process, therefore leading to higher CoT reasoning quality. We apply these pipelines to several existing datasets used in \citet{ghosh2025audio} and \citet{goel2025audio}, and the resulting CoT dataset is collectively called \texttt{AF-CoT-Train}, which includes 1.24M reasoning chains after filtering.

\subsection{Evaluation Benchmarks}
There are several broad audio understanding benchmarks that include subsets for audio reasoning evaluation, including MMAU \citep{sakshi2024mmau}, AudioBench \citep{wang2024audiobench}, AIR-Bench \citep{yang2024air}, and OpenAudioBench \citep{li2025baichuan}. These benchmarks mostly measure surface-level reasoning that may not require deep reasoning such as CoT. 

Several reasoning benchmarks focus on speech or speech plus sound \citep{wang2025they,yang2025sakura}, logical reasoning of acoustic properties or events \citep{ccoban2024mllms,deshmukh2025audio,diao2025soundmind}, temporal composition \citep{ghosh2023compa}. A more comprehensive introduction can be found in this survey \citep{yang2025towards}. Our work is orthogonal to these works.

MMAR \citep{ma2025mmar} is the most related benchmark to our proposed benchmark. It is a very challenging hand-crafted QA benchmark focused on deep reasoning that may require single- or multi-modality analysis among speech, sound, and music. 

To further facilitate the evaluation of CoT reasoning of audio understanding and complement existing benchmarks, we introduce a reasoning benchmark (\texttt{AF-Reasoning-Eval}) focused on sound understanding with two subsets. The first subset is a human curated QA benchmark with a focus on common-sense reasoning. The second subset is a deliberately constructed, challenging classification benchmark where choices are similar to each other. 

\section{Evaluation Benchmark: \texttt{AF-Reasoning-Eval}}

In this section, we introduce a new test benchmark, \texttt{AF-Reasoning-Eval}, for evaluation of sound reasoning in ALMs. It has two subsets: \texttt{AQA} for audio question answering, and \texttt{Classification} for audio classification. These test samples are designed to require reasoning to be answered.

\subsection{Audio Question Answering Subset}

The AQA subset is derived from the test split of Clotho-AQA \citep{lipping2022clotho}, where we manually annotate more challenging question-answer pairs based on the original audio and questions. 
Specifically, we first select samples with consistent three annotations -- as these samples may be have higher quality. 
We then design an interactive method similar to Alg. \ref{alg: subquestions}, where we let GPT-4 and Audio Flamingo 2 talk to each other and create the reasoning steps for these samples. We find that most of the outputs just include very simple or repetitive sub-questions, indicating the original questions are too simple. 
Therefore, we \textit{manually annotate} more challenging questions and design harder options for these samples, with a focus on reasoning with common sense (e.g. the joint occurrence of several types of sounds will exclude the possibility of a setting where only one of the sound can exist). The resulting annotated testset includes 150 samples in total, where 74 of them are binary questions (\texttt{AF-Reasoning-Eval-AQA-Yes/No}) and 76 of them are multiple choice questions (\texttt{AF-Reasoning-Eval-AQA-MCQ}). 

Below are some examples of each type of test samples. 

\begin{tcolorbox}[boxsep=1pt,left=2pt,right=2pt,top=1pt,bottom=1pt,colback=blue!5!white,colframe=gray!75!black]
\texttt{AF-Reasoning-Eval-AQA-Yes/No} samples\\
\texttt{> Is the door old?\textbackslash n(A) yes. (B) no.}\\
\texttt{> Is the cat in a good mood?\textbackslash n(A) yes. (B) no.}\\
\texttt{> Is this object likely safe for children?\textbackslash n(A) yes. (B) no.}\\
~~\\
\texttt{AF-Reasoning-Eval-AQA-MCQ} samples\\
\texttt{> Where does this activity likely happen?\textbackslash n(A) in the wild. (B) urban. (C) beach. (D) playground.}\\
\texttt{> what is not a reason of the produced sounds?\textbackslash n(A) communication. (B) echolocating. (C) attraction. (D) defense.}\\
\texttt{> Where can this sound happen?\textbackslash n(A) concert hall. (B) classroom. (C) amphitheatre. (D) church.}
\end{tcolorbox}

\subsection{Classification Subset}

The classification subset is derived from the test split of FSD50K \citep{fonseca2021fsd50k}. We leverage the annotated hierarchies in FSD50K to create classification problems where options are very similar to each other (e.g. options can belong to the same sound category such as those under bowed instruments and therefore harder to distinguish). To achieve this, we first create a graph for the annotated hierarchies and prune it into a tree structure with automatic and manual error correction. The final tree structure has 18 level-1 categories (the largest one is \texttt{music}), 120 leaf nodes, 175 total nodes, and largest depth 5. 
Then, for each test sample in FSD50K, we take the leaf node of its annotation as the ground truth, and select its nearby leaf nodes (siblings, then cousins, and so on) as other choices -- thus constructing very challenging classification problems. The detailed process is in Alg. \ref{alg: fsd50k construction}. There are 7227 test samples in total, which we call \texttt{AF-Reasoning-Eval-CLS-full}. We randomly pick 300 samples as \texttt{AF-Reasoning-Eval-CLS-mini} for ablation study and evaluation of paid baselines. 

\begin{algorithm}[!h]
\caption{Curation of \texttt{AF-Reasoning-Eval-CLS} benchmark}
\begin{algorithmic}
\State \texttt{\# Clean label taxonomy}
\State Build a weighted bidirectional graph from FSD50K annotations.
\State Remove edges such that each child has exactly one parent (the dominant parent or root). 
\State Confirm no cycles in the graph, and the graph becomes a tree.
\State Error correction: apply heuristics to remove, merge, and move improper leaves or branches.
\State Remove leaves whose parent is root. 
\State 

\State \texttt{\# Create closely related choices}
\For{node in leaves}
    \State choices $\gets$ siblings (other children of parent)
    \State If there are too few siblings, choices $\gets$ siblings and cousins (children of siblings of parent)
    \State Use LLM to remove choices that are undistinguishable from node.
\EndFor
\State 

\State \texttt{\# Create evaluation benchmark}
\For{sample in testset}
    \State ground truth $\gets$ first label
    \State other choices $\gets$ 3 random choices from closely related choices of ground truth (add random labels if no enough choices).
\EndFor

\end{algorithmic}
\label{alg: fsd50k construction}
\end{algorithm}

\section{Training dataset: \texttt{AF-CoT-Train}}

In this section, we introduce \texttt{AF-CoT-Train}, a collection of CoT training data for ALMs. There are four different data generation pipelines used in \texttt{AF-CoT-Train}, two for AQA (see Section \ref{sec: AQA pipelines}), and two for classification (see Section \ref{sec: classification pipelines}). These pipelines leverage more involved interactions between LLMs and ALMs unlike prior works (see overview in Section \ref{sec: prior CoT data overview}). We use Qwen2.5-Omni \citep{xu2025qwen2_5_omni} as the ALM and Qwen3-8B \citep{yang2025qwen3} as the LLM.

\subsection{Overview of Prior Works on CoT Data Generation}
\label{sec: prior CoT data overview}

In this section, we provide a brief overview on existing CoT data generation methods including LLaVA-CoT (for vision) and those for ALMs. We mark the use of ALM/VLM in \textcolor{purple}{\text{purple}} color, indicating where in the pipeline the multimodal query is applied.

\paragraph{LLaVA-CoT} directly queries GPT-4o on the four aspects (summary, caption, reasoning, conclusion) in its template, and apply some format checks and filtering methods to ensure high quality. The overview of the process is in Alg. \ref{alg: LLaVA-CoT}. This direct distillation method is limited by the multimodal teacher's ability and cannot leverage the stronger reasoning abilities of text-only LLMs. 

\begin{algorithm}[!h]
\caption{An overview of data generation pipeline from LLaVA-CoT \citep{xu2024llava}}
\begin{algorithmic}
\State \textbf{Inputs:} $\mathbf{X}$ (image), $\mathbf{Q}$ (question), $\mathbf{A}$ (ground truth output).
\State 

\State \texttt{\# Construct reasoning chain}
\State $\mathbf{R}\gets$  \textcolor{purple}{\text{VLM}}-reasoning$(\mathbf{Q}, \mathbf{A}, \mathbf{V})$\texttt{~~\#~structured, step-by-step}
\State

\State \texttt{\# Filtering}
\If{$\mathbf{R}$ has correct format and LLM-validate$(\mathbf{R}, \mathbf{A})$}
    \State \Return $\mathbf{R}$
\Else
    \State \Return{$\emptyset$}
\EndIf 

\end{algorithmic}
\label{alg: LLaVA-CoT}
\end{algorithm}

\paragraph{CoTA} first generates a caption for an audio, and ask a text-only LLM to generate questions, answers, and reasoning chains based on the caption. The overview is in Alg. \ref{alg: CoTA}. While this approach leverages the strong reasoning ability of the text-only LLM, it represents an audio with only a caption -- which could be noisy and details might be lost -- and therefore the generated reasoning chain could be noisy and hallucinated. 

\begin{algorithm}[!h]
\caption{An overview of data generation pipeline from CoTA \citep{xie2025audio}}
\begin{algorithmic}
\State \textbf{Inputs:} $\mathbf{X}$ (audio).
\State 

\State \texttt{\# Stage 1: label audio and create QA pairs.}
\State $\mathbf{V} \gets \textcolor{purple}{\text{ALM}}(\mathbf{X}, \text{captioning prompt})$ 
\State $\mathbf{C} \gets$ LLM-generated sub-questions and sub-answers based on $\mathbf{V}$
\State 

\State \texttt{\# Stage 2: construct structured reasoning chain}
\State $\mathbf{Q}, \mathbf{A}, \mathbf{R} \gets$ LLM-reasoning$(\mathbf{C})$ \texttt{~~\#~question, answer, and reasoning}
\State 

\State \texttt{\# Stage 3: filtering}
\If{LLM-validate$(\mathbf{X},\mathbf{V},\mathbf{Q}, \mathbf{A}, \mathbf{R})$}
    \State \Return $\mathbf{Q}, \mathbf{A}, \mathbf{R}$
\Else
    \State \Return{$\emptyset$}
\EndIf 

\end{algorithmic}
\label{alg: CoTA}
\end{algorithm}

\paragraph{SARI} is very similar to CoTA and has more steps on QA quality filtering at the beginning. The overview is in Alg. \ref{alg: SARI}. However, the same problems in CoTA exist in SARI too.

\begin{algorithm}[!h]
\caption{An overview of data generation pipeline from SARI \citep{wen2025sari}}
\begin{algorithmic}
\State \textbf{Inputs:} $\mathbf{X}$ (audio), $\mathbf{V}$ (optional caption).
\State 

\State \texttt{\# Stage 1: label audio and create QA pairs.}
\If{$\mathbf{V}=\emptyset$}
    \State $\mathbf{V} \gets \textcolor{purple}{\text{ALM}}(\mathbf{X}, \text{captioning prompt})$ 
\EndIf
\State $\mathbf{Q}, \mathbf{A} \gets$ LLM-generated QA based on $\mathbf{V}$
\If{$\textcolor{purple}{\text{ALM}}(\mathbf{X}, \mathbf{Q})\neq \mathbf{A}$ for all 16 attempts}
    \State \Return{$\emptyset$}
\EndIf
\State

\State \texttt{\# Stage 2: construct reasoning chain}
\State $\mathbf{R}\gets$  LLM-reasoning$(\mathbf{Q}, \mathbf{A}, \mathbf{V})$\texttt{~~\#~structured or unstructured}
\State 

\State \texttt{\# Stage 3: filtering}
\If{LLM-validate$(\mathbf{R}, \mathbf{A})$}
    \State \Return $\mathbf{Q}, \mathbf{A}, \mathbf{R}$
\Else
    \State \Return{$\emptyset$}
\EndIf 

\end{algorithmic}
\label{alg: SARI}
\end{algorithm}

\paragraph{AudSem} improves CoTA by using more metadata than just audio caption. The metadata are extracted from both ALMs (on audio) and VLMs (on video), and therefore contain much more information. The overview of the process is in Alg. \ref{alg: AudSem}. While this could reduce audio-specific hallucination, its stage-2 reasoning process still focuses on the reasoning in the text domain, which excludes audio-specific reasoning. In addition, the usefulness of extracted vision metadata cannot be guaranteed as some of the metadata is unrelated to the audio, and in certain cases the video itself is not related to audio (e.g. edited video). 

\begin{algorithm}[!h]
\caption{An overview of data generation pipeline from AudSem \citep{wijngaard2025audsemthinker}}
\begin{algorithmic}
\State \textbf{Inputs:} $\mathbf{X}$ (audio with video).
\State 

\State \texttt{\# Stage 1: label audio and video.}
\State $\mathbf{V}_{\mathrm{audio}} \gets \textcolor{purple}{\text{ALMs}}(\mathbf{X})$ 
\State $\mathbf{V}_{\mathrm{vision}} \gets \textcolor{purple}{\text{VLMs}}(\mathbf{X})$ 
\State $\mathbf{V} \gets \mathbf{V}_{\mathrm{audio}} \cup \mathbf{V}_{\mathrm{vision}}$
\State 

\State \texttt{\# Stage 2: construct reasoning chain with semantic descriptors}
\State $\mathbf{Q}, \mathbf{A}, \mathbf{R} \gets$ LLM-reasoning$(\mathbf{V})$ \texttt{~~\#~question, answer, and reasoning}
\State 

\State \texttt{\# Stage 3: filtering}
\If{LLM-validate$(\mathbf{X},\mathbf{Q}, \mathbf{A}, \mathbf{R})$}
    \State \Return $\mathbf{Q}, \mathbf{A}, \mathbf{R}$
\Else
    \State \Return{$\emptyset$}
\EndIf 

\end{algorithmic}
\label{alg: AudSem}
\end{algorithm}

In summary, the existing ALM CoT data generation methods mostly rely on applying a text-only LLM to reason over a text representation of audio (caption or more metadata). This limits the audio-specific reasoning abilities. We aim to close this gap by applying ALMs in the reasoning step rather than just using it to obtain a text representation of audio. 

\subsection{CoT Data Generation Pipeline for AQA}
\label{sec: AQA pipelines}

We propose two pipelines for AQA CoT data generation. The first pipeline is based on parallel sub-questions. In this pipeline, an LLM first breaks the question into many small sub-questions. Next, an ALM answers each sub-question. Then, an LLM validates if the generated QAs lead to the ground truth. If the validation is passed, an LLM rephrases generated QAs into the LLaVA-CoT template. This approach is analogous to the \textit{BFS-style} search. It can be implemented in parallel and therefore faster, and this pipeline has high acceptance rates during validation. However, it may be hard to generate very deep reasoning chains with this pipeline. The process is illustrated in Alg. \ref{alg: subquestions}. \footnote{We use $[\cdot]$ to represent a list, $+$ to represent concatenation of lists, and $\{\cdot\}$ to represent a tuple.}

\begin{algorithm}[!h]
\caption{Create step-by-step training data through extensive parallel sub-questions}
\begin{algorithmic}
\State \textbf{Inputs:} $\mathbf{X}$ (audio), $\mathbf{Q}$ (question), $\mathbf{A}$ (ground truth output), $T$ (min length of reasoning chain).
\State 

\State \texttt{\# Create sub-questions}
\State $\mathbf{S} \gets$ LLM-generated sub-questions for $\mathbf{Q}$

\If{length of $\mathbf{S}<T$}
    \State \Return{$\emptyset$}
\EndIf
\State 
\State \texttt{\# Create reasoning chain}
\State $\mathbf{S} \gets [\text{captioning prompt}] + \mathbf{S}$

\State $\mathbf{C} \gets []$

\For{$\mathbf{s}\in\mathbf{S}$}
    \State $\mathbf{C} \gets \mathbf{C} + [\{\mathbf{s}, \textcolor{purple}{\text{ALM}}(\mathbf{X}, \mathbf{s})\}]$
\EndFor
\State 

\State \texttt{\# Filtering}
\If{LLM-predict$(\mathbf{Q};\mathbf{C})=\mathbf{A}$ or LLM-validate$(\mathbf{Q},\mathbf{A};\mathbf{C})$}
    \State $\mathbf{R}\gets$ LLM-rephrase$(\mathbf{C})$
    \State \Return $\mathbf{R}$
\Else
    \State \Return{$\emptyset$}
\EndIf 

\end{algorithmic}
\label{alg: subquestions}
\end{algorithm}

The second pipeline is based on interactive conversations. In this pipeline, an LLM and an ALM talk with each other for multiple rounds. In each round, the LLM generates a new suitable question based on existing generated QAs, and the ALM answers that question. The conversation ends when the LLM can make a confident prediction of the original complex question. If the prediction is correct, the validation is passed, and an LLM rephrases the conversation into the LLaVA-CoT template. This approach is analogous to the \textit{DFS-style} search. This pipeline may be able to generate deeper reasoning chains. However, the reasoning chain needs to be generated round-by-round rather than in parallel, and the rejection rate is higher. The process is illustrated in Alg. \ref{alg: interactive}.

\begin{algorithm}[!h]
\caption{Create step-by-step training data through interactive conversation}
\begin{algorithmic}
\State \textbf{Inputs:} $\mathbf{X}$ (audio), $\mathbf{Q}$ (question), $\mathbf{A}$ (ground truth output), $T$ (min length of reasoning chain).
\State 

\State \texttt{\# Initialize reasoning chain}
\State $\mathbf{S} \gets [\text{captioning prompt}]$
\State $\mathbf{C} \gets [\{\mathbf{S}_{-1}, \textcolor{purple}{\text{ALM}}(\mathbf{X}, \mathbf{S}_{-1})\}]$
\State 

\State \texttt{\# Create reasoning chain}
\While{LLM-predict$(\mathbf{Q}; \mathbf{C})$ is \texttt{unsure}}
    \State $\mathbf{S} \gets \mathbf{S} + [\text{LLM-next-step}(\mathbf{Q};\mathbf{C})]$
    \State $\mathbf{C} \gets \mathbf{C} + [\{\mathbf{S}_{-1}, \textcolor{purple}{\text{ALM}}(\mathbf{X}, \mathbf{S}_{-1})\}]$
\EndWhile
\State 

\State \texttt{\# Filtering}
\If{length of $\mathbf{C}< T+1$}
    \State \Return{$\emptyset$}
\ElsIf{LLM-predict$(\mathbf{Q};\mathbf{C})=\mathbf{A}$ or LLM-validate$(\mathbf{Q},\mathbf{A};\mathbf{C})$}
    \State $\mathbf{R}\gets$ LLM-rephrase$(\mathbf{C})$
    \State \Return $\mathbf{R}$
\Else
    \State \Return{$\emptyset$}
\EndIf 

\end{algorithmic}
\label{alg: interactive}
\end{algorithm}

In summary, we apply these two pipelines to AudioSkills \citep{ghosh2025audio} and Clotho-AQA \citep{lipping2022clotho}. We obtain 811K CoT data for close-ended AQA and 306K CoT data for open-ended AQA. The majority of them are focused on sound understanding.

\subsection{CoT Data Generation Pipeline for Classification}
\label{sec: classification pipelines}

We propose two pipelines for classification CoT data generation. The first pipeline is designed for multiple choice questions. In detail, for each choice, we ask an LLM to generate its acoustic descriptions, and then ask an ALM whether the acoustic properties fit the audio. We apply the similar validation and rephrasing steps as in the previous section. The process is illustrated in Alg. \ref{alg: MCQ comparison}.

We apply this pipeline to several classification datasets including Chime-Home \citep{foster2015chime}, CochlScene \citep{jeong2022cochlscene}, ESC \citep{piczak2015dataset}, FMA \citep{defferrard2016fma}, FSD50K \citep{fonseca2021fsd50k}, GTZAN \citep{tzanetakis2002musical}, Medley-solos-DB \citep{lostanlen_2019_1344103}, NonSpeech7K \citep{rashid2023nonspeech7k}, TUT \citep{DCASE2017challenge}, and US8K \citep{salamon2014dataset}. We obtain about 120K CoT data in total.  

\begin{algorithm}[!t]
\caption{Create step-by-step training data for sound classification with choices}
\begin{algorithmic}
\State \textbf{Inputs:} $\mathbf{X}$ (audio), $\mathbf{Q}$ (classification prompt with choices), $\mathbf{A}$ (ground truth output).
\State 

\State \texttt{\# Initialize reasoning chain}
\State $\mathbf{S} \gets [\text{captioning prompt}]$
\State $\mathbf{C} \gets [\{\mathbf{S}_{-1}, \textcolor{purple}{\text{ALM}}(\mathbf{X}, \mathbf{S}_{-1})\}]$
\State 

\State \texttt{\# Create reasoning chain}
\For{$\texttt{choice}\in\mathbf{Q}$}
    \State $\mathbf{S} \gets \mathbf{S} + [\text{describe acoustic properties of \texttt{choice}}, \text{verify if sound fits \texttt{choice}}]$
    \State $\mathbf{C} \gets \mathbf{C} + [\{\mathbf{S}_{-2}, \text{LLM}(\mathbf{S}_{-2})\}, \{\mathbf{S}_{-1}, \textcolor{purple}{\text{ALM}}(\mathbf{X}, \mathbf{S}_{-1})\}]$
\EndFor
\State

\State \texttt{\# Filtering}
\If{LLM-predict$(\mathbf{Q};\mathbf{C})=\mathbf{A}$ or LLM-validate$(\mathbf{Q},\mathbf{A};\mathbf{C})$}
    \State $\mathbf{R}\gets$ LLM-rephrase$(\mathbf{C})$
    \State \Return $\mathbf{R}$
\Else
    \State \Return{$\emptyset$}
\EndIf

\end{algorithmic}
\label{alg: MCQ comparison}
\end{algorithm}

The second pipeline is designed for direct classification without choices. We leverage the hierarchical annotations from FSD50K and concatenate a series of reasoning chains, one for each hierarchy. At each hierarchy layer, we construct multiple choice questions using sibling nodes, and create the reasoning chains similar to the previous pipeline. This turns a complex zero-shot classification task into simpler sequential classification tasks. We summarize the process in Alg. \ref{alg: hierarchy}.

\begin{algorithm}[!t]
\caption{Create step-by-step training data through sound hierarchy}
\begin{algorithmic}
\State \textbf{Inputs:} $\mathbf{X}$ (audio), $\mathbf{Q}$ (classification prompt), $\mathbf{A}$ (ground truth output), $\mathbf{H}$ (sound hierarchy).
\State 

\State \texttt{\# Initialize reasoning chain}
\State $\mathbf{S} \gets [\text{captioning prompt}]$
\State $\mathbf{C} \gets [\{\mathbf{S}_{-1}, \textcolor{purple}{\text{ALM}}(\mathbf{X}, \mathbf{S}_{-1})\}]$
\State 

\State \texttt{\# Create reasoning chain}
\For{$\mathbf{h}\in\mathbf{H}$}
    \State $\hat{\mathbf{h}} \gets \mathrm{shuffle}([\mathbf{h}] + \text{siblings}(\mathbf{h})$)
    \For{$\texttt{choice}\in\hat{\mathbf{h}}$}
        \State $\mathbf{S} \gets \mathbf{S} + [\text{describe acoustic properties of \texttt{choice}}, \text{verify if sound fits \texttt{choice}}]$
        \State $\mathbf{C} \gets \mathbf{C} + [\{\mathbf{S}_{-2}, \text{LLM}(\mathbf{S}_{-2})\}, \{\mathbf{S}_{-1}, \textcolor{purple}{\text{ALM}}(\mathbf{X}, \mathbf{S}_{-1})\}]$
    \EndFor
    \State

    \State \texttt{\# Filtering at each step}
    \State $\hat{\mathbf{Q}} \gets \text{format-question}(\mathbf{Q}, \hat{\mathbf{h}})$
    \If{LLM-predict$(\hat{\mathbf{Q}};\mathbf{C})\neq\mathbf{h}$ and $\neg$ LLM-validate$(\hat{\mathbf{Q}},\mathbf{h};\mathbf{C})$}
        \State \Return{$\emptyset$}
    \EndIf 

\EndFor
\State 

\State \texttt{\# Filtering}
\If{LLM-validate$(\mathbf{Q},\mathbf{A};\mathbf{C})$}
    \State $\mathbf{R}\gets$ LLM-rephrase$(\mathbf{C})$
    \State \Return $\mathbf{R}$
\Else
    \State \Return{$\emptyset$}
\EndIf

\end{algorithmic}
\label{alg: hierarchy}
\end{algorithm}

\section{Experiments}

We finetune Audio Flamingo 2 \citep{ghosh2025audio} and Audio Flamingo 3 \citep{goel2025audio} on the \texttt{AF-CoT-Train} dataset to verify the effectiveness of this CoT dataset on sound reasoning benchmarks. For Audio Flamingo 2 we resume from the stage-2 model, and for Audio Flamingo 3 we resume from the stage-3 model. During finetuning, we add \texttt{AF-CoT-Train} to the SFT dataset of the corresponding stage, and remove the original non-CoT samples that were used to construct \texttt{AF-CoT-Train}. \footnote{Removing these samples turns out to be quite beneficial in preliminary experiments.} 

We name our finetuned models Audio Flamingo 2 Sound-CoT and Audio Flamingo 3 Sound-CoT, respectively. We evaluate our models and compare to baselines on four sound reasoning benchmarks: our \texttt{AF-Reasoning-Eval (AQA)}, our \texttt{AF-Reasoning-Eval (Classification)}, \texttt{MMAR-Sound} (the sound subset of \texttt{MMAR}) \citep{ma2025mmar}, and \texttt{MMAU-Sound} (the sound subset of \texttt{MMAU-v05.15.25}) \citep{sakshi2024mmau}.

\subsection{Main Results}

\paragraph{\texttt{AF-Reasoning-Eval}}
The main results for the AQA subset are included in Table \ref{tab: main results ClothoAQA}. The improvements on Audio Flamingo 2 are very significant, making it comparable to several 7B reasoning baselines. We do not see improvements on Audio Flamingo 3 on this benchmark, but the results are similar. We find Qwen2.5-Omni to be very strong on this benchmark; however, its training data were not disclosed so we cannot evaluate if there was potential data leakage.

The main results for the classification subset are included in Table \ref{tab: main results FSD50K}. The improvements to both Audio Flamingo 2 and 3 are significant, as expected. This is likely because the classification subset of \texttt{AF-CoT-Train} includes many complex classification samples with closely related options. 

\begin{table}[!h]
\centering
\caption{
Main results on our proposed \texttt{AF-Reasoning-Eval (AQA)} benchmark (abbreviated as \texttt{AFR-Eval-AQA} in this table). ${\dag}$: open-sourced large audio models with reasoning abilities. $\clubsuit$: audio training data not disclosed. $\diamondsuit$: closed-source models.
\colorbox{pink}{Best number} denotes the best result across open-source models.
\colorbox{pink!50}{Second best number} denotes the runner-up across open-source models.
The best result across all open and closed-source models are in \textbf{bold fonts}.
}
\fontsize{8}{10}\selectfont
\begin{tabular}{llcc}

\toprule
\textbf{Models} & \textbf{Base model} & \textbf{\texttt{AFR-Eval-AQA-Yes/No}} & \textbf{\texttt{AFR-Eval-AQA-MCQ}} \\ 
\midrule
Aero-1 Audio & Qwen2.5 (\textbf{1.5B}) & 48.65 & 48.68 \\
Phi-4 Multimodal$^{\clubsuit}$   & Phi‑4‑Mini Instruct (\textbf{5.6B})   & 66.22 & 36.84 \\ 
Audio-Reasoner$^{\dag}$ & Qwen2-Audio Instruct (\textbf{7B}) & 71.62 & 61.84 \\
Kimi Audio & Qwen2.5 (\textbf{7B}) & 78.38 & 61.84 \\
Qwen2.5-Omni$^{\clubsuit}$ & Qwen2.5 (\textbf{7B}) & \bestnumber{\textbf{88.16}} & \bestnumber{\textbf{81.58}} \\
AudSemThinker$^{\dag}$ & Qwen2.5-Omni$^{\clubsuit}$ (\textbf{7B}) & 81.08 & 71.05 \\
\cdashline{1-4}
\noalign{\vskip 0.4mm}
GPT-4o Audio$^{\clubsuit\diamondsuit}$ & N/A & 75.68 & 71.05 \\
Gemini-1.5-pro$^{\clubsuit\diamondsuit}$    & N/A & 70.27 & 56.58 \\
Gemini-2.5-flash$^{\clubsuit\diamondsuit}$  & N/A & 72.97 & 75.00 \\
Gemini-2.5-pro$^{\clubsuit\diamondsuit}$    & N/A & 83.78 & 67.11    \\ 
\midrule
Audio Flamingo 2  & Qwen2.5 (\textbf{3B}) & 71.62 & 42.11 \\ 
Audio Flamingo 2 Sound-CoT  & Audio Flamingo 2 (\textbf{3B}) & \runnerup{83.78} & 64.47 \\
\cdashline{1-4}
\noalign{\vskip 0.4mm}
Audio Flamingo 3  & Qwen2.5 (\textbf{7B}) & 81.08 & \runnerup{75.00} \\ 
Audio Flamingo 3 Sound-CoT  & Audio Flamingo 3 (\textbf{7B}) & 79.73 & 73.68 \\
\bottomrule
\end{tabular}
\label{tab: main results ClothoAQA}
\end{table}

\begin{table}[!h]
\centering
\caption{
Main results on our proposed \texttt{AF-Reasoning-Eval (Classification)} benchmark (abbreviated as \texttt{AFR-Eval-CLS} in this table). ${\dag}$: open-sourced large audio models with reasoning abilities. $\clubsuit$: audio training data not disclosed. $\diamondsuit$: closed-source models.
\colorbox{pink}{Best number} denotes the best result across open-source models.
\colorbox{pink!50}{Second best number} denotes the runner-up across open-source models.
The best result across all open and closed-source models are in \textbf{bold fonts}.
}
\fontsize{8}{10}\selectfont
\begin{tabular}{llcc}

\toprule
\textbf{Models} & \textbf{Base model} & \textbf{\texttt{AFR-Eval-CLS-mini}} & \textbf{\texttt{AFR-Eval-CLS-full}} \\ 
\midrule
AF-CLAP & N/A & 59.67 & 56.99 \\
\cdashline{1-4}
\noalign{\vskip 0.4mm}
Aero-1 Audio & Qwen2.5 (\textbf{1.5B}) & 48.67 & 48.90 \\
Phi-4 Multimodal$^{\clubsuit}$   & Phi‑4‑Mini Instruct (\textbf{5.6B})   & 29.00 & 32.09 \\ 
Audio-Reasoner$^{\dag}$ & Qwen2-Audio Instruct (\textbf{7B}) & 63.67 & 64.99 \\
Qwen2.5-Omni$^{\clubsuit}$ & Qwen2.5 (\textbf{7B}) & 64.67 & 62.97 \\
AudSemThinker$^{\dag}$ & Qwen2.5-Omni$^{\clubsuit}$ (\textbf{7B}) & 69.67 & 73.75 \\
\cdashline{1-4}
\noalign{\vskip 0.4mm}
GPT-4o Audio$^{\clubsuit\diamondsuit}$ & N/A & 68.00 & - \\
Gemini-2.5-flash$^{\clubsuit\diamondsuit}$  & N/A & 73.67 & - \\
Gemini-2.5-pro$^{\clubsuit\diamondsuit}$    & N/A & 76.67 & -    \\ 
\midrule
Audio Flamingo 2  & Qwen2.5 (\textbf{3B}) & 42.67 & 41.52 \\
Audio Flamingo 2 Sound-CoT  & Audio Flamingo 2 (\textbf{3B}) & \runnerup{80.67} & \runnerup{82.45} \\
\cdashline{1-4}
\noalign{\vskip 0.4mm}
Audio Flamingo 3  & Qwen2.5 (\textbf{7B}) & 75.33 & 74.07 \\
Audio Flamingo 3 Sound-CoT  & Audio Flamingo 3 (\textbf{7B}) & \bestnumber{\textbf{90.67}} & \bestnumber{\textbf{88.85}} \\
\bottomrule
\end{tabular}
\label{tab: main results FSD50K}
\end{table}

\paragraph{\texttt{MMAR-Sound}} The main results on the sound subset of \texttt{MMAR} are shown in Table \ref{tab: main results MMAR}. There is a $6.6\%$ improvement to Audio Flamingo 2 and a $3.6\%$ improvement to Audio Flamingo 3. The closed source models, especially Gemini-2.0-flash, is the state-of-the-art on this task, possibly because it has better native reasoning ability in the text domain. 

\begin{table}[!h]
\centering
\caption{
Main results on \texttt{MMAR} (sound subset). ${\dag}$: open-sourced large audio models with reasoning abilities. $\clubsuit$: audio training data not disclosed. $\diamondsuit$: closed-source models.
\colorbox{pink}{Best number} denotes the best result across open-source models.
\colorbox{pink!50}{Second best number} denotes the runner-up across open-source models.
The best result across all open and closed-source models are in \textbf{bold fonts}.
}
\fontsize{8}{10}\selectfont
\begin{tabular}{llc}

\toprule
\textbf{Models} & \textbf{Base model} & \texttt{MMAR-Sound} \\ 
\midrule
LTU                 &     LLaMA (\textbf{7B}) & 19.39 \\
GAMA-IT             &    LLaMA-2 (\textbf{7B}) & 22.42 \\
Qwen2-Audio Instruct&    Qwen2-Audio (\textbf{7B}) & 33.33 \\
Audio-CoT$^{\dag}$ &  Qwen2-Audio Instruct (\textbf{7B}) & 35.76 \\
Audio-Reasoner$^{\dag}$ & Qwen2-Audio Instruct (\textbf{7B}) & 43.64 \\
Baichuan-Omni-1.5$^{\clubsuit}$ & Qwen2.5 (\textbf{7B}) & 41.21 \\
Qwen2.5-Omni$^{\clubsuit}$ & Qwen2.5 (\textbf{7B}) & \bestnumber{58.79} \\
Salmonn             &    Vicuna (\textbf{13B}) & 30.30 \\
\cdashline{1-3}
\noalign{\vskip 0.4mm}
GPT-4o mini Audio$^{\clubsuit\diamondsuit}$  & N/A & 38.79 \\
GPT-4o Audio$^{\clubsuit\diamondsuit}$ & N/A & 53.94 \\
Gemini-2.0-flash$^{\clubsuit\diamondsuit}$    & N/A & \textbf{61.21} \\ 
\midrule
Audio Flamingo 2  &    Qwen2.5 (\textbf{3B}) & 49.09 \\
Audio Flamingo 2 Sound-CoT  & Audio Flamingo 2 (\textbf{3B}) & 55.76 \\
\cdashline{1-3}
\noalign{\vskip 0.4mm}
Audio Flamingo 3  & Qwen2.5 (\textbf{7B}) & 53.33 \\
Audio Flamingo 3 Sound-CoT  & Audio Flamingo 3 (\textbf{7B}) & \runnerup{56.97} \\
\bottomrule
\end{tabular}
\label{tab: main results MMAR}
\end{table}

\paragraph{\texttt{MMAU-Sound}} The main results on the sound subset of \texttt{MMAU-v05.15.25} are shown in Table \ref{tab: main results MMAU}. There is about $7\%$ improvement on Audio Flamingo 2 and $4\%$ improvement on Audio Flamingo 3. Audio Flamingo 2 Sound-CoT is close to the strongest 7B models, and Audio Flamingo 3 Sound-CoT achieves the state-of-the-art on this subset, demonstrating the effectiveness of the proposed \texttt{AF-CoT-Train}. 

\begin{table}[!h]
\centering
\caption{
Main results on \texttt{MMAU-v05.15.25} (sound subset). ${\dag}$: open-sourced large audio models with reasoning abilities. $\clubsuit$: audio training data not disclosed. $\diamondsuit$: closed-source models.
\colorbox{pink}{Best number} denotes the best result across open-source models.
\colorbox{pink!50}{Second best number} denotes the runner-up across open-source models.
The best result across all open and closed-source models are in \textbf{bold fonts}.
}
\fontsize{8}{10}\selectfont
\begin{tabular}{llc}

\toprule
\textbf{Models} & \textbf{Base model} & \texttt{MMAU-v05.15.25-Sound} \\ \midrule 
Phi-4 Multimodal $^{\clubsuit}$   & Phi-4-Mini Instruct (\textbf{5.6B})   & 62.67 \\ 
LTU             &  LLaMA (\textbf{7B})    & 20.67 \\ 
GAMA-IT         & LLaMA-2 (\textbf{7B})    & 32.73 \\ 
Qwen2-Audio Instruct    & Qwen2-Audio (\textbf{7B})     & 61.17 \\ 
Audio-Reasoner$^{\dag}$  & Qwen2-Audio Instruct (\textbf{7B}) & 67.27 \\ 
Kimi Audio      & Qwen2.5 (\textbf{7B})    & 70.70 \\ 
Qwen2.5-Omni$^{\clubsuit}$   & Qwen2.5 (\textbf{7B})    & \runnerup{76.77} \\ 
Salmonn         &  Vicuna (\textbf{13B})   & 42.10 \\ 
\cdashline{1-3}
\noalign{\vskip 0.4mm}
GPT-4o mini Audio$^{\clubsuit\diamondsuit}$ & N/A & 49.67 \\
GPT-4o Audio$^{\clubsuit\diamondsuit}$      & N/A & 63.20 \\
Gemini-2.5-flash-lite$^{\clubsuit\diamondsuit}$    & N/A & 62.50 \\
Gemini-2.5-flash$^{\clubsuit\diamondsuit}$  & N/A & 69.50 \\
Gemini-2.5-pro$^{\clubsuit\diamondsuit}$    & N/A & 70.63 \\

\midrule
Audio Flamingo 2            & Qwen2.5 (\textbf{3B})    & 68.13 \\
Audio Flamingo 2 Sound-CoT   & Audio Flamingo 2 (\textbf{3B})   & 75.23 \\ 
\cdashline{1-3}
\noalign{\vskip 0.4mm}
Audio Flamingo 3            & Qwen2.5 (\textbf{7B})    & {75.83} \\
Audio Flamingo 3 Sound-CoT  & Audio Flamingo 3 (\textbf{7B}) & \bestnumber{\textbf{79.83}} \\
\bottomrule
\end{tabular}
\label{tab: main results MMAU}
\end{table}

\subsection{Ablation Studies}

In this section, we conduct ablation studies on batch size, data blending, and CoT generation method to understand the effect of each of them. All the ablation experiments are conducted using Audio Flamingo 2, where the reference model is the Audio Flamingo 2 Sound-CoT reported in the main results. 

\paragraph{Effect of batch size.} The ablation study on batch size is shown in Table \ref{tab: ablation batch size}. The effect of batch size is mixed across benchmarks: MMAU and \texttt{AF-Reasoning-Eval-CLS} favor smaller batch sizes lead to overall worse accuracies, but MMAR and \texttt{AF-Reasoning-Eval-AQA} favor larger batch sizes.

\begin{table}[!h]
\centering
\caption{Ablation study on batch size.}
\fontsize{8}{10}\selectfont
\begin{tabular}{lccccc}

\toprule
\multirow{2}{*}{\textbf{Batch size}} & \multicolumn{3}{c}{\textbf{\texttt{AF-Reasoning-Eval}}} & \textbf{\texttt{MMAR}} & \textbf{\texttt{MMAU}} \\ 
\cmidrule(lr){2-4}
\cmidrule(lr){5-5}
\cmidrule(lr){6-6}
& {\textbf{\texttt{AQA-Yes/No}}} & {\textbf{\texttt{AQA-MCQ}}} & {\textbf{\texttt{CLS-mini}}} & \textbf{Sound} & \textbf{Sound-mini} \\ \midrule 
Reference (512)        & \textbf{83.78} & \textbf{64.47} & 80.67 & \textbf{55.76} & 75.98 \\
Reference$/2$ (256)    & 78.39 & 56.59 & \textbf{85.33} & 49.06 & 77.18 \\
Reference$/4$ (128)    & 74.32 & 60.53 & 83.67 & 50.30 & \textbf{78.98} \\

\bottomrule
\end{tabular}
\label{tab: ablation batch size}
\end{table}

\paragraph{Data blending.} The ablation study on data blending of non-CoT data is shown in Table \ref{tab: ablation data blending}. We find it helpful to moderately reduce the percentage of non-CoT data, but just finetuning on CoT data only may lead to overall accuracy drop. 

\begin{table}[!h]
\centering
\caption{Ablation study on data blending.}
\fontsize{8}{10}\selectfont
\begin{tabular}{lccccc}

\toprule
\multirow{2}{*}{\textbf{Non-CoT Data}} & \multicolumn{3}{c}{\textbf{\texttt{AF-Reasoning-Eval}}} & \textbf{\texttt{MMAR}} & \textbf{\texttt{MMAU}} \\ 
\cmidrule(lr){2-4}
\cmidrule(lr){5-5}
\cmidrule(lr){6-6}
& {\textbf{\texttt{AQA-Yes/No}}} & {\textbf{\texttt{AQA-MCQ}}} & {\textbf{\texttt{CLS-mini}}} & \textbf{Sound} & \textbf{Sound-mini} \\ \midrule 
Reference           & \textbf{83.78} & \textbf{64.47} & \textbf{80.67} & \textbf{55.76} & 75.98 \\
More non-CoT data   & 77.02 & 55.26 & 79.00 & 50.91 & 74.17 \\
CoT data only   & 74.32 & 53.95 & 79.33 & 52.73 & \textbf{76.88} \\

\bottomrule
\end{tabular}
\label{tab: ablation data blending}
\end{table}

\paragraph{Sub-questions vs interactive conversations.} The ablation study on the CoT generation methods for AQA is shown in Table \ref{tab: ablation CoT method}. We find, surprisingly, the \textit{BFS-style} sub-question method (Alg. \ref{alg: subquestions}) is considerably better than the \textit{DFS-style} interactive conversation method (Alg. \ref{alg: interactive}). This indicates that the lack of ultra deep reasoning in the former method does not harm current benchmarks. This further confirms (1) the quality of the latter has room for improvement, as also evidenced by its lower acceptance rate, and (2) current benchmarks can be tackled without the need of very deep reasoning, but instead require broad reasoning from different aspects. 

\begin{table}[!h]
\centering
\caption{Ablation study on CoT generation method.}
\fontsize{8}{10}\selectfont
\begin{tabular}{lcccc}

\toprule
\multirow{2}{*}{\textbf{CoT generation method}} & \multicolumn{2}{c}{\textbf{\texttt{AF-Reasoning-Eval}}} & \textbf{\texttt{MMAR}} & \textbf{\texttt{MMAU}} \\ 
\cmidrule(lr){2-3}
\cmidrule(lr){4-4}
\cmidrule(lr){5-5}
& {\textbf{\texttt{AQA-Yes/No}}} & {\textbf{\texttt{AQA-MCQ}}} & \textbf{Sound} & \textbf{Sound-mini} \\ \midrule 
Reference (Alg. \ref{alg: subquestions} only) & \textbf{83.78} & \textbf{64.47} & \textbf{55.76} & \textbf{75.98} \\
Alg. \ref{alg: interactive} only & 72.97 & 52.63 & 44.85 & 75.08 \\
Alg. \ref{alg: subquestions} +  Alg. \ref{alg: interactive}  & 72.97 & 59.21 & 50.30 & 72.67 \\

\bottomrule
\end{tabular}
\label{tab: ablation CoT method}
\end{table}

\subsection{Discussion}

\paragraph{Measuring causality and reasoning quality.} 
In order to understand the reasoning quality and causality (whether the predicted answer is based on reasoning outputs), we manually checked 300 outputs of our two Audio Flamingo Sound-CoT models on \texttt{AF-Reasoning-Eval-AQA}, and decompose the correct and wrong predictions into multiple categories defined by reasoning correctness and causality for investigation. The results are presented in Table \ref{tab: ablation reasoning}. There are several findings from the reported numbers. 

\begin{enumerate}
    \item Audio Flamingo 3 has better causality than Audio Flamingo 2 (over $10\%$), likely because Audio Flamingo 3 itself is the latest and therefore better at instruction following. If we assume the model makes uniformly random guesses when there is no causality, then the de-biased causality is about $0.76\sim0.78$ for Audio Flamingo 3 Sound-CoT and $0.56\sim0.61$ for Audio Flamingo 2 Sound-CoT. These numbers indicate that there is still considerable room for causality improvements, and RL might be a better tool to solve this problem. 

    \item There is also considerable room for reasoning accuracy improvements, which we will address in our future work. To better evaluate the reasoning accuracies and to prepare for RL training, this calls the importance of a good reward model that could examine the reasoning outputs including accuracies and hallucinations. 

    \item The first row of the "correct" section (where both reasoning is correct and causality is true) is what we would like to optimize. There is a consistent $10\%$ improvements from Audio Flamingo 2 to 3, which is expected due to the larger base model and better data used in the latter model.

    \item The second row of the "correct" and "wrong" sections (where reasoning is wrong and causality is true) represent typical errors where reasoning is wrong and therefore the prediction is wrong, which could be reduced by having better reasoning itself.

    \item The third row of the "correct" section and the first row of the "wrong" section (where reasoning is wrong and causality is false) represent cases where the model hallucinates on reasoning and makes prediction without looking at the reasoning outputs. Especially, in Audio Flamingo 2 there are $20\%$ of such cases for binary questions and they happen to lead to correct predictions -- leading to higher overall accuracies. These cases are not desirable and likely the most challenging to mitigate.

    \item The third row of the "wrong" section (where reasoning is correct but causality is false) are close to being correct -- if the model chooses to look at the reasoning outputs. We expect to mitigate these failure cases by applying RL on causality.
\end{enumerate}

\begin{table}[!h]
\centering
\caption{Measuring causality and reasoning quality with human evaluation. Reasoning correctness is determined by whether the reasoning outputs between \texttt{<reasoning>} and \texttt{</reasoning>} are correct for the audio and the question. Causality is determined by whether the predicted answer between \texttt{<conclusion>} and \texttt{</conclusion>} follows the reasoning outputs.}
\fontsize{8}{10}\selectfont
\begin{tabular}{lcccccc}

\toprule

\multicolumn{3}{c}{\textbf{Benchmark}} & \multicolumn{2}{c}{\texttt{AF-Reasoning-Eval-AQA-Yes/No}} & \multicolumn{2}{c}{\texttt{AF-Reasoning-Eval-AQA-MCQ}} \\ 
\cmidrule(lr){1-3}
\cmidrule(lr){4-5}
\cmidrule(lr){6-7}
\multicolumn{3}{c}{\textbf{Model}} & {\fontsize{6}{8}\selectfont AF2 Sound-CoT} & {\fontsize{6}{8}\selectfont AF3 Sound-CoT} & {\fontsize{6}{8}\selectfont AF2 Sound-CoT} & {\fontsize{6}{8}\selectfont AF3 Sound-CoT} \\ \midrule
\multirow{4}{*}{\textbf{Correct}} & \textbf{Reasoning} & \textbf{Causality} & & & & \\
\cmidrule(lr){2-3}
& \cmark & \cmark & 60.81 & 71.62 & 52.63 & 63.16 \\
& \xmark & \cmark & 2.70 & 5.41 & 3.95 & 2.63 \\
& \xmark & \xmark & 20.27 & 2.70 & 7.89 & 7.89 \\ \midrule
\multirow{4}{*}{\textbf{Wrong}} & \textbf{Reasoning} & \textbf{Causality} & & & & \\
\cmidrule(lr){2-3}
& \xmark & \xmark & 0.00 & 1.35 & 10.53 & 7.89 \\
& \xmark & \cmark & 14.86 & 12.16 & 14.47 & 15.79 \\
& \cmark & \xmark & 1.35 & 6.76 & 10.53 & 2.63 \\
\midrule 
\multicolumn{3}{l}{\textbf{Prediction accuracy}} & 83.78 & 79.73 & 64.47 & 73.68 \\
\multicolumn{3}{l}{\textbf{Reasoning accuracy}} & 62.16 & 78.38 & 63.16 & 65.79 \\
\multicolumn{3}{l}{\textbf{Causality}} & 78.38 & 89.19 & 71.05 & 81.58 \\
\bottomrule
\end{tabular}
\label{tab: ablation reasoning}
\end{table}

\paragraph{Speech and Music Reasoning.}
We do not observe statistically significant changes in speech and music reasoning accuracies after finetuning on mostly sound-related CoT data. This is likely because we included a small fraction of speech-sound and music data in \texttt{AF-CoT-Train}, but the quantity is not enough to improve the results. We also believe that building CoT data for speech and music understanding requires more experts (e.g. timestamped speech recognition models, music foundation models, better LLMs), which is an important future work. 

\paragraph{Marginal Gains.}
We observe marginal gains of finetuning Audio Flamingo 3 (7B) on \texttt{AF-CoT-Train} compared to gains on Audio Flamingo 2 (3B). This is likely because Audio Flamingo 3 is larger and much better at these benchmarks already. While further scaling up \texttt{AF-CoT-Train} might be useful. The results indicate that applying RL is likely a more efficient and effective way to improve larger and more capable models like Audio Flamingo 3.

\section{Conclusion}

This technical report aims to advance audio understanding with chain-of-thought (CoT) reasoning so that these models not only achieve higher benchmark scores but also become more transparent. We propose a benchmark \texttt{AF-Reasoning-Eval} for (1) sound AQA evaluation with a focus on common sense reasoning and (2) sound classification evaluation with a focus on discriminating closely related options. We propose CoT training dataset \texttt{AF-CoT-Train} by applying four CoT data generation pipelines to a number of existing datasets. We finetune Audio Flamingo 2 and 3 on this dataset and observe improvements on several benchmarks, confirming the effectiveness of this CoT dataset. With extensive ablation studies, we identify the optimal recipe and decompose the failure patterns of current models to guide our future research.

There are a number of questions not answered in this technical report, which we will address in our future work. First, it is unclear whether supervised fine-tuning or RL could offer more gains, or we might need both. Given that our improvements on Audio Flamingo 2 is more significant than Audio Flamingo 3, we hypothesize that supervised fine-tuning might be good for smaller and weaker models, while RL is necessary for larger and more powerful models. Second, it is unclear how to better evaluate the quality of reasoning outputs, which is both an important filtering tool in data curation and a potential reward model in RL. Third, it is unclear what is the best recipe to construct CoT training data and blend it with existing data, especially in the current complicated data setting where there are numerous data sources and tasks with highly various quantity, quality, and difficulty.

\newpage
\bibliography{main.bib}
\bibliographystyle{iclr2025_conference}

\end{document}

%% file: preamble.tex
\usepackage[colorlinks=true,citecolor=brown,urlcolor=gray]{hyperref}
\usepackage[skip=3pt]{caption}
\usepackage{algorithm}
\usepackage{algpseudocode}

\usepackage{tcolorbox}
\usepackage{pgfplots,pgfplotstable}
\usepgfplotslibrary{groupplots}
\usepackage{pifont}

\newcommand{\cmark}{\textcolor{teal}{\ding{51}}} 
\newcommand{\xmark}{\textcolor{red}{\ding{55}}} 

\usepackage{bbm} 

\usepackage{siunitx}
\usepackage{colortbl}
\newcommand{\bestnumber}[1]{\cellcolor{pink}{#1}}
\newcommand{\runnerup}[1]{\cellcolor{pink!50}{#1}}

\usepackage{amsmath,amsfonts}
\usepackage{microtype}
\usepackage{graphicx}
\usepackage{subcaption}
\usepackage{booktabs} 

\usepackage{xcolor}

\usepackage{multirow}
\usepackage{amssymb}

\usepackage[T1]{fontenc}      
\usepackage{lmodern}          
\usepackage{arydshln}

\usepackage{hyperref}
\usepackage{url}

\usepackage[capitalize,noabbrev]{cleveref}

\definecolor{forestgreen}{rgb}{0.13, 0.55, 0.13}
\definecolor{fulvous}{rgb}{0.86, 0.52, 0.0}
\definecolor{glaucous}{rgb}{0.38, 0.51, 0.71}
\definecolor{lava}{rgb}{0.81, 0.06, 0.13}
\definecolor{buff}{rgb}{0.94, 0.86, 0.51}
\definecolor{chromeyellow}{rgb}{1.0, 0.65, 0.0}
\definecolor{brightube}{rgb}{0.82, 0.62, 0.91}

\definecolor{CustomBlue}{RGB}{76, 114, 176}
\definecolor{CustomRed}{RGB}{196, 78, 82}

\usepackage{interval}
\usepackage{tabularx}
\usepackage{color}
\usepackage{xspace}
\usepackage{mathtools}  
\usepackage{siunitx}    

\usepackage{attachfile2}
\newcommand{\audiofile}[2][]{%
  \textattachfile[
      mimetype=audio/mpeg,
      appearance=true,
      #1
  ]{#2}{\includegraphics[width=1em]{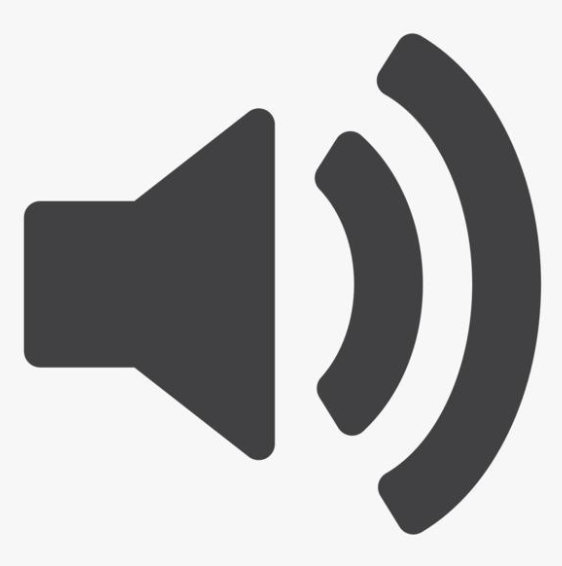}}
}

\usepackage{placeins}
\usepackage{algorithm}








%% file: main.bbl
\begin{thebibliography}{68}
\providecommand{\natexlab}[1]{#1}
\providecommand{\url}[1]{\texttt{#1}}
\expandafter\ifx\csname urlstyle\endcsname\relax
  \providecommand{\doi}[1]{doi: #1}\else
  \providecommand{\doi}{doi: \begingroup \urlstyle{rm}\Url}\fi

\bibitem[Abouelenin et~al.(2025)Abouelenin, Ashfaq, Atkinson, Awadalla, Bach,
  Bao, Benhaim, Cai, Chaudhary, Chen, et~al.]{abouelenin2025phi}
Abdelrahman Abouelenin, Atabak Ashfaq, Adam Atkinson, Hany Awadalla, Nguyen
  Bach, Jianmin Bao, Alon Benhaim, Martin Cai, Vishrav Chaudhary, Congcong
  Chen, et~al.
\newblock Phi-4-mini technical report: Compact yet powerful multimodal language
  models via mixture-of-loras.
\newblock \emph{arXiv preprint arXiv:2503.01743}, 2025.

\bibitem[Alayrac et~al.(2022)Alayrac, Donahue, Luc, Miech, Barr, Hasson, Lenc,
  Mensch, Millican, Reynolds, et~al.]{alayrac2022flamingo}
Jean-Baptiste Alayrac, Jeff Donahue, Pauline Luc, Antoine Miech, Iain Barr,
  Yana Hasson, Karel Lenc, Arthur Mensch, Katherine Millican, Malcolm Reynolds,
  et~al.
\newblock Flamingo: a visual language model for few-shot learning.
\newblock \emph{Advances in neural information processing systems},
  35:\penalty0 23716--23736, 2022.

\bibitem[Azzolini et~al.(2025)Azzolini, Bai, Brandon, Cao, Chattopadhyay, Chen,
  Chu, Cui, Diamond, Ding, et~al.]{azzolini2025cosmos}
Alisson Azzolini, Junjie Bai, Hannah Brandon, Jiaxin Cao, Prithvijit
  Chattopadhyay, Huayu Chen, Jinju Chu, Yin Cui, Jenna Diamond, Yifan Ding,
  et~al.
\newblock Cosmos-reason1: From physical common sense to embodied reasoning.
\newblock \emph{arXiv preprint arXiv:2503.15558}, 2025.

\bibitem[Bercovich et~al.(2025)Bercovich, Levy, Golan, Dabbah, El-Yaniv, Puny,
  Galil, Moshe, Ronen, Nabwani, et~al.]{bercovich2025llama}
Akhiad Bercovich, Itay Levy, Izik Golan, Mohammad Dabbah, Ran El-Yaniv, Omri
  Puny, Ido Galil, Zach Moshe, Tomer Ronen, Najeeb Nabwani, et~al.
\newblock Llama-nemotron: Efficient reasoning models.
\newblock \emph{arXiv preprint arXiv:2505.00949}, 2025.

\bibitem[Chen et~al.(2024{\natexlab{a}})Chen, Qin, Zhang, Chen, Xu, and
  Che]{chen2024m}
Qiguang Chen, Libo Qin, Jin Zhang, Zhi Chen, Xiao Xu, and Wanxiang Che.
\newblock M$^3$cot: A novel benchmark for multi-domain multi-step multi-modal
  chain-of-thought.
\newblock \emph{arXiv preprint arXiv:2405.16473}, 2024{\natexlab{a}}.

\bibitem[Chen et~al.(2023)Chen, Sikka, Cogswell, Ji, and
  Divakaran]{chen2023measuring}
Yangyi Chen, Karan Sikka, Michael Cogswell, Heng Ji, and Ajay Divakaran.
\newblock Measuring and improving chain-of-thought reasoning in vision-language
  models.
\newblock \emph{arXiv preprint arXiv:2309.04461}, 2023.

\bibitem[Chen et~al.(2024{\natexlab{b}})Chen, Zhou, Shen, Hong, Sun, Gutfreund,
  and Gan]{chen2024visual}
Zhenfang Chen, Qinhong Zhou, Yikang Shen, Yining Hong, Zhiqing Sun, Dan
  Gutfreund, and Chuang Gan.
\newblock Visual chain-of-thought prompting for knowledge-based visual
  reasoning.
\newblock In \emph{Proceedings of the AAAI Conference on Artificial
  Intelligence}, volume~38, pp.\  1254--1262, 2024{\natexlab{b}}.

\bibitem[Chu et~al.(2023)Chu, Xu, Zhou, Yang, Zhang, Yan, Zhou, and
  Zhou]{chu2023qwen}
Yunfei Chu, Jin Xu, Xiaohuan Zhou, Qian Yang, Shiliang Zhang, Zhijie Yan, Chang
  Zhou, and Jingren Zhou.
\newblock Qwen-audio: Advancing universal audio understanding via unified
  large-scale audio-language models.
\newblock \emph{arXiv preprint arXiv:2311.07919}, 2023.

\bibitem[Chu et~al.(2024)Chu, Xu, Yang, Wei, Wei, Guo, Leng, Lv, He, Lin,
  et~al.]{chu2024qwen2}
Yunfei Chu, Jin Xu, Qian Yang, Haojie Wei, Xipin Wei, Zhifang Guo, Yichong
  Leng, Yuanjun Lv, Jinzheng He, Junyang Lin, et~al.
\newblock Qwen2-audio technical report.
\newblock \emph{arXiv preprint arXiv:2407.10759}, 2024.

\bibitem[{\c{C}}oban et~al.(2024){\c{C}}oban, Mandel, and
  Devaney]{ccoban2024mllms}
Enis~Berk {\c{C}}oban, Michael~I Mandel, and Johanna Devaney.
\newblock What do mllms hear? examining the interaction between llm and audio
  encoder components in multimodal large language models.
\newblock In \emph{Audio Imagination: NeurIPS 2024 Workshop AI-Driven Speech,
  Music, and Sound Generation}, 2024.

\bibitem[Comanici et~al.(2025)Comanici, Bieber, Schaekermann, Pasupat,
  Sachdeva, Dhillon, Blistein, Ram, Zhang, Rosen, et~al.]{comanici2025gemini}
Gheorghe Comanici, Eric Bieber, Mike Schaekermann, Ice Pasupat, Noveen
  Sachdeva, Inderjit Dhillon, Marcel Blistein, Ori Ram, Dan Zhang, Evan Rosen,
  et~al.
\newblock Gemini 2.5: Pushing the frontier with advanced reasoning,
  multimodality, long context, and next generation agentic capabilities.
\newblock \emph{arXiv preprint arXiv:2507.06261}, 2025.

\bibitem[Dang et~al.(2025)Dang, Gao, and Jia]{dang2025scaling}
Ting Dang, Yan Gao, and Hong Jia.
\newblock Scaling auditory cognition via test-time compute in audio language
  models.
\newblock \emph{arXiv preprint arXiv:2503.23395}, 2025.

\bibitem[Defferrard et~al.(2016)Defferrard, Benzi, Vandergheynst, and
  Bresson]{defferrard2016fma}
Micha{\"e}l Defferrard, Kirell Benzi, Pierre Vandergheynst, and Xavier Bresson.
\newblock Fma: A dataset for music analysis.
\newblock \emph{arXiv preprint arXiv:1612.01840}, 2016.

\bibitem[Deng et~al.(2023)Deng, Ma, Liu, Guo, Zhang, Chen, Huang, and
  Benetos]{deng2023musilingo}
Zihao Deng, Yinghao Ma, Yudong Liu, Rongchen Guo, Ge~Zhang, Wenhu Chen, Wenhao
  Huang, and Emmanouil Benetos.
\newblock Musilingo: Bridging music and text with pre-trained language models
  for music captioning and query response.
\newblock \emph{arXiv preprint arXiv:2309.08730}, 2023.

\bibitem[Deshmukh et~al.(2023)Deshmukh, Elizalde, Singh, and
  Wang]{deshmukh2023pengi}
Soham Deshmukh, Benjamin Elizalde, Rita Singh, and Huaming Wang.
\newblock Pengi: An audio language model for audio tasks.
\newblock \emph{Advances in Neural Information Processing Systems},
  36:\penalty0 18090--18108, 2023.

\bibitem[Deshmukh et~al.(2025{\natexlab{a}})Deshmukh, Dixit, Singh, and
  Raj]{deshmukh2025mellow}
Soham Deshmukh, Satvik Dixit, Rita Singh, and Bhiksha Raj.
\newblock Mellow: a small audio language model for reasoning.
\newblock \emph{arXiv preprint arXiv:2503.08540}, 2025{\natexlab{a}}.

\bibitem[Deshmukh et~al.(2025{\natexlab{b}})Deshmukh, Han, Bukhari, Elizalde,
  Gamper, Singh, and Raj]{deshmukh2025audio}
Soham Deshmukh, Shuo Han, Hazim Bukhari, Benjamin Elizalde, Hannes Gamper, Rita
  Singh, and Bhiksha Raj.
\newblock Audio entailment: Assessing deductive reasoning for audio
  understanding.
\newblock In \emph{Proceedings of the AAAI Conference on Artificial
  Intelligence}, volume~39, pp.\  23769--23777, 2025{\natexlab{b}}.

\bibitem[Diao et~al.(2025)Diao, Zhang, Kong, Wu, Ma, Ouyang, Qing, Vosoughi,
  and Gui]{diao2025soundmind}
Xingjian Diao, Chunhui Zhang, Keyi Kong, Weiyi Wu, Chiyu Ma, Zhongyu Ouyang,
  Peijun Qing, Soroush Vosoughi, and Jiang Gui.
\newblock Soundmind: Rl-incentivized logic reasoning for audio-language models.
\newblock \emph{arXiv preprint arXiv:2506.12935}, 2025.

\bibitem[Ding et~al.(2025)Ding, Ju, Leng, Liu, Liu, Shang, Shen, Song, Tan,
  Tang, et~al.]{ding2025kimi}
Ding Ding, Zeqian Ju, Yichong Leng, Songxiang Liu, Tong Liu, Zeyu Shang, Kai
  Shen, Wei Song, Xu~Tan, Heyi Tang, et~al.
\newblock Kimi-audio technical report.
\newblock \emph{arXiv preprint arXiv:2504.18425}, 2025.

\bibitem[Fonseca et~al.(2021)Fonseca, Favory, Pons, Font, and
  Serra]{fonseca2021fsd50k}
Eduardo Fonseca, Xavier Favory, Jordi Pons, Frederic Font, and Xavier Serra.
\newblock Fsd50k: an open dataset of human-labeled sound events.
\newblock \emph{IEEE/ACM Transactions on Audio, Speech, and Language
  Processing}, 30:\penalty0 829--852, 2021.

\bibitem[Foster et~al.(2015)Foster, Sigtia, Krstulovic, Barker, and
  Plumbley]{foster2015chime}
Peter Foster, Siddharth Sigtia, Sacha Krstulovic, Jon Barker, and Mark~D
  Plumbley.
\newblock Chime-home: A dataset for sound source recognition in a domestic
  environment.
\newblock In \emph{2015 IEEE Workshop on Applications of Signal Processing to
  Audio and Acoustics (WASPAA)}, pp.\  1--5. IEEE, 2015.

\bibitem[Ghosh et~al.(2023)Ghosh, Seth, Kumar, Tyagi, Evuru, Ramaneswaran,
  Sakshi, Nieto, Duraiswami, and Manocha]{ghosh2023compa}
Sreyan Ghosh, Ashish Seth, Sonal Kumar, Utkarsh Tyagi, Chandra~Kiran Evuru,
  S~Ramaneswaran, S~Sakshi, Oriol Nieto, Ramani Duraiswami, and Dinesh Manocha.
\newblock Compa: Addressing the gap in compositional reasoning in
  audio-language models.
\newblock \emph{arXiv preprint arXiv:2310.08753}, 2023.

\bibitem[Ghosh et~al.(2024)Ghosh, Kumar, Seth, Evuru, Tyagi, Sakshi, Nieto,
  Duraiswami, and Manocha]{ghosh2024gama}
Sreyan Ghosh, Sonal Kumar, Ashish Seth, Chandra Kiran~Reddy Evuru, Utkarsh
  Tyagi, S~Sakshi, Oriol Nieto, Ramani Duraiswami, and Dinesh Manocha.
\newblock Gama: A large audio-language model with advanced audio understanding
  and complex reasoning abilities.
\newblock \emph{arXiv preprint arXiv:2406.11768}, 2024.

\bibitem[Ghosh et~al.(2025)Ghosh, Kong, Kumar, Sakshi, Kim, Ping, Valle,
  Manocha, and Catanzaro]{ghosh2025audio}
Sreyan Ghosh, Zhifeng Kong, Sonal Kumar, S~Sakshi, Jaehyeon Kim, Wei Ping,
  Rafael Valle, Dinesh Manocha, and Bryan Catanzaro.
\newblock Audio flamingo 2: An audio-language model with long-audio
  understanding and expert reasoning abilities.
\newblock In \emph{Forty-second International Conference on Machine Learning},
  2025.
\newblock URL \url{https://openreview.net/forum?id=xWu5qpDK6U}.

\bibitem[Goel et~al.(2025)Goel, Ghosh, Kim, Kumar, Kong, Lee, Yang, Duraiswami,
  Manocha, Valle, and Catanzaro]{goel2025audio}
Arushi Goel, Sreyan Ghosh, Jaehyeon Kim, Sonal Kumar, Zhifeng Kong, Sang-gil
  Lee, Chao-Han~Huck Yang, Ramani Duraiswami, Dinesh Manocha, Rafael Valle, and
  Bryan Catanzaro.
\newblock Audio flamingo 3: Advancing audio intelligence with fully open large
  audio language models.
\newblock \emph{arXiv preprint arXiv:2507.08128}, 2025.

\bibitem[Gong et~al.(2023{\natexlab{a}})Gong, Liu, Luo, Karlinsky, and
  Glass]{gong2023joint}
Yuan Gong, Alexander~H Liu, Hongyin Luo, Leonid Karlinsky, and James Glass.
\newblock Joint audio and speech understanding.
\newblock In \emph{2023 IEEE Automatic Speech Recognition and Understanding
  Workshop (ASRU)}, pp.\  1--8. IEEE, 2023{\natexlab{a}}.

\bibitem[Gong et~al.(2023{\natexlab{b}})Gong, Luo, Liu, Karlinsky, and
  Glass]{gong2023listen}
Yuan Gong, Hongyin Luo, Alexander~H Liu, Leonid Karlinsky, and James Glass.
\newblock Listen, think, and understand.
\newblock \emph{arXiv preprint arXiv:2305.10790}, 2023{\natexlab{b}}.

\bibitem[Guo et~al.(2025)Guo, Yang, Zhang, Song, Zhang, Xu, Zhu, Ma, Wang, Bi,
  et~al.]{guo2025deepseek}
Daya Guo, Dejian Yang, Haowei Zhang, Junxiao Song, Ruoyu Zhang, Runxin Xu,
  Qihao Zhu, Shirong Ma, Peiyi Wang, Xiao Bi, et~al.
\newblock Deepseek-r1: Incentivizing reasoning capability in llms via
  reinforcement learning.
\newblock \emph{arXiv preprint arXiv:2501.12948}, 2025.

\bibitem[Hurst et~al.(2024)Hurst, Lerer, Goucher, Perelman, Ramesh, Clark,
  Ostrow, Welihinda, Hayes, Radford, et~al.]{hurst2024gpt}
Aaron Hurst, Adam Lerer, Adam~P Goucher, Adam Perelman, Aditya Ramesh, Aidan
  Clark, AJ~Ostrow, Akila Welihinda, Alan Hayes, Alec Radford, et~al.
\newblock Gpt-4o system card.
\newblock \emph{arXiv preprint arXiv:2410.21276}, 2024.

\bibitem[Jeong \& Park(2022)Jeong and Park]{jeong2022cochlscene}
Il-Young Jeong and Jeongsoo Park.
\newblock Cochlscene: Acquisition of acoustic scene data using crowdsourcing.
\newblock In \emph{2022 Asia-Pacific Signal and Information Processing
  Association Annual Summit and Conference (APSIPA ASC)}, pp.\  17--21. IEEE,
  2022.

\bibitem[Kong et~al.(2024)Kong, Goel, Badlani, Ping, Valle, and
  Catanzaro]{kong2024audio}
Zhifeng Kong, Arushi Goel, Rohan Badlani, Wei Ping, Rafael Valle, and Bryan
  Catanzaro.
\newblock Audio flamingo: A novel audio language model with few-shot learning
  and dialogue abilities.
\newblock In \emph{International Conference on Machine Learning}, pp.\
  25125--25148. PMLR, 2024.

\bibitem[Li et~al.(2025{\natexlab{a}})Li, Loy, Fanyi, Jingkang, Kaichen,
  Kairui, Minh, Quang, Ba, Shuai, Yezhen, and Ziwei]{li2025aero}
Bo~Li, Chen~Change Loy, Pu~Fanyi, Yang Jingkang, Zhang Kaichen, Hu~Kairui,
  Thang~Luu Minh, Trung~Nguyen Quang, Cong~Pham Ba, Liu Shuai, Wang Yezhen, and
  Liu Ziwei.
\newblock Aero: Audio-enhanced large language models.
\newblock 2025{\natexlab{a}}.
\newblock URL \url{https://www.lmms-lab.com/posts/aero_audio/}.

\bibitem[Li et~al.(2025{\natexlab{b}})Li, Liu, Zhang, Fang, Pan, Wang, Liang,
  Li, Lin, Dong, et~al.]{li2025baichuan}
Tianpeng Li, Jun Liu, Tao Zhang, Yuanbo Fang, Da~Pan, Mingrui Wang, Zheng
  Liang, Zehuan Li, Mingan Lin, Guosheng Dong, et~al.
\newblock Baichuan-audio: A unified framework for end-to-end speech
  interaction.
\newblock \emph{arXiv preprint arXiv:2502.17239}, 2025{\natexlab{b}}.

\bibitem[Li et~al.(2025{\natexlab{c}})Li, Liu, Zhang, Chen, Li, Li, Liu, Ming,
  Dong, Pan, et~al.]{li2025baichuanomni}
Yadong Li, Jun Liu, Tao Zhang, Song Chen, Tianpeng Li, Zehuan Li, Lijun Liu,
  Lingfeng Ming, Guosheng Dong, Da~Pan, et~al.
\newblock Baichuan-omni-1.5 technical report.
\newblock \emph{arXiv preprint arXiv:2501.15368}, 2025{\natexlab{c}}.

\bibitem[Lipping et~al.(2022)Lipping, Sudarsanam, Drossos, and
  Virtanen]{lipping2022clotho}
Samuel Lipping, Parthasaarathy Sudarsanam, Konstantinos Drossos, and Tuomas
  Virtanen.
\newblock Clotho-aqa: A crowdsourced dataset for audio question answering.
\newblock In \emph{2022 30th European Signal Processing Conference (EUSIPCO)},
  pp.\  1140--1144. IEEE, 2022.

\bibitem[Liu et~al.(2023)Liu, Li, Wu, and Lee]{liu2023visual}
Haotian Liu, Chunyuan Li, Qingyang Wu, and Yong~Jae Lee.
\newblock Visual instruction tuning.
\newblock \emph{Advances in neural information processing systems},
  36:\penalty0 34892--34916, 2023.

\bibitem[Liu et~al.(2024)Liu, Hussain, Sun, and Shan]{liu2024music}
Shansong Liu, Atin~Sakkeer Hussain, Chenshuo Sun, and Ying Shan.
\newblock Music understanding llama: Advancing text-to-music generation with
  question answering and captioning.
\newblock In \emph{ICASSP 2024-2024 IEEE International Conference on Acoustics,
  Speech and Signal Processing (ICASSP)}, pp.\  286--290. IEEE, 2024.

\bibitem[Liu et~al.(2025)Liu, Yang, Chen, Lee, Shoeybi, Catanzaro, and
  Ping]{liu2025acereason}
Zihan Liu, Zhuolin Yang, Yang Chen, Chankyu Lee, Mohammad Shoeybi, Bryan
  Catanzaro, and Wei Ping.
\newblock Acereason-nemotron 1.1: Advancing math and code reasoning through sft
  and rl synergy.
\newblock \emph{arXiv preprint arXiv:2506.13284}, 2025.

\bibitem[Lostanlen et~al.(2019)Lostanlen, Cella, Bittner, and
  Essid]{lostanlen_2019_1344103}
Vincent Lostanlen, Carmine-Emanuele Cella, Rachel Bittner, and Slim Essid.
\newblock {Medley-solos-DB: a cross-collection dataset for musical instrument
  recognition}, February 2019.
\newblock URL \url{https://doi.org/10.5281/zenodo.1344103}.

\bibitem[Ma et~al.(2025{\natexlab{a}})Ma, Chen, Wang, Chng, and
  Chen]{ma2025audio}
Ziyang Ma, Zhuo Chen, Yuping Wang, Eng~Siong Chng, and Xie Chen.
\newblock Audio-cot: Exploring chain-of-thought reasoning in large audio
  language model.
\newblock \emph{arXiv preprint arXiv:2501.07246}, 2025{\natexlab{a}}.

\bibitem[Ma et~al.(2025{\natexlab{b}})Ma, Ma, Zhu, Yang, Chao, Xu, Chen, Chen,
  Chen, Cong, et~al.]{ma2025mmar}
Ziyang Ma, Yinghao Ma, Yanqiao Zhu, Chen Yang, Yi-Wen Chao, Ruiyang Xu, Wenxi
  Chen, Yuanzhe Chen, Zhuo Chen, Jian Cong, et~al.
\newblock Mmar: A challenging benchmark for deep reasoning in speech, audio,
  music, and their mix.
\newblock \emph{arXiv preprint arXiv:2505.13032}, 2025{\natexlab{b}}.

\bibitem[Mesaros et~al.(2017)Mesaros, Heittola, Diment, Elizalde, Shah,
  Vincent, Raj, and Virtanen]{DCASE2017challenge}
A.~Mesaros, T.~Heittola, A.~Diment, B.~Elizalde, A.~Shah, E.~Vincent, B.~Raj,
  and T.~Virtanen.
\newblock {DCASE} 2017 challenge setup: Tasks, datasets and baseline system.
\newblock In \emph{Proceedings of the Detection and Classification of Acoustic
  Scenes and Events 2017 Workshop (DCASE2017)}, pp.\  85--92, November 2017.

\bibitem[Piczak()]{piczak2015dataset}
Karol~J. Piczak.
\newblock {ESC}: {Dataset} for {Environmental Sound Classification}.
\newblock In \emph{Proceedings of the 23rd {Annual ACM Conference} on
  {Multimedia}}, pp.\  1015--1018. {ACM Press}.
\newblock ISBN 978-1-4503-3459-4.
\newblock \doi{10.1145/2733373.2806390}.
\newblock URL \url{http://dl.acm.org/citation.cfm?doid=2733373.2806390}.

\bibitem[Rashid et~al.(2023)Rashid, Li, and Du]{rashid2023nonspeech7k}
Muhammad~Mamunur Rashid, Guiqing Li, and Chengrui Du.
\newblock Nonspeech7k dataset: Classification and analysis of human non-speech
  sound.
\newblock \emph{IET Signal Processing}, 17\penalty0 (6):\penalty0 e12233, 2023.

\bibitem[Sakshi et~al.(2024)Sakshi, Tyagi, Kumar, Seth, Selvakumar, Nieto,
  Duraiswami, Ghosh, and Manocha]{sakshi2024mmau}
S~Sakshi, Utkarsh Tyagi, Sonal Kumar, Ashish Seth, Ramaneswaran Selvakumar,
  Oriol Nieto, Ramani Duraiswami, Sreyan Ghosh, and Dinesh Manocha.
\newblock Mmau: A massive multi-task audio understanding and reasoning
  benchmark.
\newblock \emph{arXiv preprint arXiv:2410.19168}, 2024.

\bibitem[Salamon et~al.(2014)Salamon, Jacoby, and Bello]{salamon2014dataset}
Justin Salamon, Christopher Jacoby, and Juan~Pablo Bello.
\newblock A dataset and taxonomy for urban sound research.
\newblock In \emph{Proceedings of the 22nd ACM international conference on
  Multimedia}, pp.\  1041--1044, 2014.

\bibitem[Shao et~al.(2024)Shao, Qian, Xiao, Song, Zong, Wang, Liu, and
  Li]{shao2024visual}
Hao Shao, Shengju Qian, Han Xiao, Guanglu Song, Zhuofan Zong, Letian Wang,
  Yu~Liu, and Hongsheng Li.
\newblock Visual cot: Advancing multi-modal language models with a
  comprehensive dataset and benchmark for chain-of-thought reasoning.
\newblock \emph{Advances in Neural Information Processing Systems},
  37:\penalty0 8612--8642, 2024.

\bibitem[Tang et~al.(2023)Tang, Yu, Sun, Chen, Tan, Li, Lu, Ma, and
  Zhang]{tang2023salmonn}
Changli Tang, Wenyi Yu, Guangzhi Sun, Xianzhao Chen, Tian Tan, Wei Li, Lu~Lu,
  Zejun Ma, and Chao Zhang.
\newblock Salmonn: Towards generic hearing abilities for large language models.
\newblock \emph{arXiv preprint arXiv:2310.13289}, 2023.

\bibitem[Team et~al.(2023)Team, Anil, Borgeaud, Alayrac, Yu, Soricut,
  Schalkwyk, Dai, Hauth, Millican, et~al.]{team2023gemini}
Gemini Team, Rohan Anil, Sebastian Borgeaud, Jean-Baptiste Alayrac, Jiahui Yu,
  Radu Soricut, Johan Schalkwyk, Andrew~M Dai, Anja Hauth, Katie Millican,
  et~al.
\newblock Gemini: a family of highly capable multimodal models.
\newblock \emph{arXiv preprint arXiv:2312.11805}, 2023.

\bibitem[Team et~al.(2024)Team, Georgiev, Lei, Burnell, Bai, Gulati, Tanzer,
  Vincent, Pan, Wang, et~al.]{team2024gemini}
Gemini Team, Petko Georgiev, Ving~Ian Lei, Ryan Burnell, Libin Bai, Anmol
  Gulati, Garrett Tanzer, Damien Vincent, Zhufeng Pan, Shibo Wang, et~al.
\newblock Gemini 1.5: Unlocking multimodal understanding across millions of
  tokens of context.
\newblock \emph{arXiv preprint arXiv:2403.05530}, 2024.

\bibitem[Tzanetakis \& Cook(2002)Tzanetakis and Cook]{tzanetakis2002musical}
George Tzanetakis and Perry Cook.
\newblock Musical genre classification of audio signals.
\newblock \emph{IEEE Transactions on speech and audio processing}, 10\penalty0
  (5):\penalty0 293--302, 2002.

\bibitem[Wang et~al.(2025{\natexlab{a}})Wang, Zou, Lin, Sun, Liu, Zhang, Liu,
  Aw, and Chen]{wang2024audiobench}
Bin Wang, Xunlong Zou, Geyu Lin, Shuo Sun, Zhuohan Liu, Wenyu Zhang, Zhengyuan
  Liu, AiTi Aw, and Nancy~F Chen.
\newblock Audiobench: A universal benchmark for audio large language models.
\newblock \emph{NAACL}, 2025{\natexlab{a}}.

\bibitem[Wang et~al.(2022)Wang, Wei, Schuurmans, Le, Chi, Narang, Chowdhery,
  and Zhou]{wang2022self}
Xuezhi Wang, Jason Wei, Dale Schuurmans, Quoc Le, Ed~Chi, Sharan Narang,
  Aakanksha Chowdhery, and Denny Zhou.
\newblock Self-consistency improves chain of thought reasoning in language
  models.
\newblock \emph{arXiv preprint arXiv:2203.11171}, 2022.

\bibitem[Wang et~al.(2025{\natexlab{b}})Wang, Mousavi, Ploujnikov, and
  Ravanelli]{wang2025they}
Yingzhi Wang, Pooneh Mousavi, Artem Ploujnikov, and Mirco Ravanelli.
\newblock What are they doing? joint audio-speech co-reasoning.
\newblock In \emph{ICASSP 2025-2025 IEEE International Conference on Acoustics,
  Speech and Signal Processing (ICASSP)}, pp.\  1--5. IEEE, 2025{\natexlab{b}}.

\bibitem[Wei et~al.(2022)Wei, Wang, Schuurmans, Bosma, Xia, Chi, Le, Zhou,
  et~al.]{wei2022chain}
Jason Wei, Xuezhi Wang, Dale Schuurmans, Maarten Bosma, Fei Xia, Ed~Chi, Quoc~V
  Le, Denny Zhou, et~al.
\newblock Chain-of-thought prompting elicits reasoning in large language
  models.
\newblock \emph{Advances in neural information processing systems},
  35:\penalty0 24824--24837, 2022.

\bibitem[Wen et~al.(2025)Wen, Guo, Zhao, Zou, and Li]{wen2025sari}
Cheng Wen, Tingwei Guo, Shuaijiang Zhao, Wei Zou, and Xiangang Li.
\newblock Sari: Structured audio reasoning via curriculum-guided reinforcement
  learning.
\newblock \emph{arXiv preprint arXiv:2504.15900}, 2025.

\bibitem[Wijngaard et~al.(2025)Wijngaard, Formisano, Esposito, and
  Dumontier]{wijngaard2025audsemthinker}
Gijs Wijngaard, Elia Formisano, Michele Esposito, and Michel Dumontier.
\newblock Audsemthinker: Enhancing audio-language models through reasoning over
  semantics of sound.
\newblock \emph{arXiv preprint arXiv:2505.14142}, 2025.

\bibitem[Xie et~al.(2025)Xie, Lin, Liu, Wu, Yan, and Miao]{xie2025audio}
Zhifei Xie, Mingbao Lin, Zihang Liu, Pengcheng Wu, Shuicheng Yan, and Chunyan
  Miao.
\newblock Audio-reasoner: Improving reasoning capability in large audio
  language models.
\newblock \emph{arXiv preprint arXiv:2503.02318}, 2025.

\bibitem[Xu et~al.(2024)Xu, Jin, Hao, Song, Sun, and Yuan]{xu2024llava}
Guowei Xu, Peng Jin, Li~Hao, Yibing Song, Lichao Sun, and Li~Yuan.
\newblock Llava-cot: Let vision language models reason step-by-step.
\newblock \emph{arXiv preprint arXiv:2411.10440}, 2024.

\bibitem[Xu et~al.(2025)Xu, Guo, He, Hu, He, Bai, Chen, Wang, Fan, Dang,
  et~al.]{xu2025qwen2_5_omni}
Jin Xu, Zhifang Guo, Jinzheng He, Hangrui Hu, Ting He, Shuai Bai, Keqin Chen,
  Jialin Wang, Yang Fan, Kai Dang, et~al.
\newblock Qwen2. 5-omni technical report.
\newblock \emph{arXiv preprint arXiv:2503.20215}, 2025.

\bibitem[Yang et~al.(2025{\natexlab{a}})Yang, Li, Yang, Zhang, Hui, Zheng, Yu,
  Gao, Huang, Lv, et~al.]{yang2025qwen3}
An~Yang, Anfeng Li, Baosong Yang, Beichen Zhang, Binyuan Hui, Bo~Zheng, Bowen
  Yu, Chang Gao, Chengen Huang, Chenxu Lv, et~al.
\newblock Qwen3 technical report.
\newblock \emph{arXiv preprint arXiv:2505.09388}, 2025{\natexlab{a}}.

\bibitem[Yang et~al.(2025{\natexlab{b}})Yang, Ho, Piao, and
  Lee]{yang2025sakura}
Chih-Kai Yang, Neo Ho, Yen-Ting Piao, and Hung-yi Lee.
\newblock Sakura: On the multi-hop reasoning of large audio-language models
  based on speech and audio information.
\newblock \emph{arXiv preprint arXiv:2505.13237}, 2025{\natexlab{b}}.

\bibitem[Yang et~al.(2025{\natexlab{c}})Yang, Ho, and Lee]{yang2025towards}
Chih-Kai Yang, Neo~S Ho, and Hung-yi Lee.
\newblock Towards holistic evaluation of large audio-language models: A
  comprehensive survey.
\newblock \emph{arXiv preprint arXiv:2505.15957}, 2025{\natexlab{c}}.

\bibitem[Yang et~al.(2024)Yang, Xu, Liu, Chu, Jiang, Zhou, Leng, Lv, Zhao,
  Zhou, et~al.]{yang2024air}
Qian Yang, Jin Xu, Wenrui Liu, Yunfei Chu, Ziyue Jiang, Xiaohuan Zhou, Yichong
  Leng, Yuanjun Lv, Zhou Zhao, Chang Zhou, et~al.
\newblock Air-bench: Benchmarking large audio-language models via generative
  comprehension.
\newblock \emph{arXiv preprint arXiv:2402.07729}, 2024.

\bibitem[Yao et~al.(2023)Yao, Yu, Zhao, Shafran, Griffiths, Cao, and
  Narasimhan]{yao2023tree}
Shunyu Yao, Dian Yu, Jeffrey Zhao, Izhak Shafran, Tom Griffiths, Yuan Cao, and
  Karthik Narasimhan.
\newblock Tree of thoughts: Deliberate problem solving with large language
  models.
\newblock \emph{Advances in neural information processing systems},
  36:\penalty0 11809--11822, 2023.

\bibitem[Zhang et~al.(2024)Zhang, Zhang, Li, Zhang, Sun, Gan, Yang, Pang, and
  Yang]{zhang2024improve}
Ruohong Zhang, Bowen Zhang, Yanghao Li, Haotian Zhang, Zhiqing Sun, Zhe Gan,
  Yinfei Yang, Ruoming Pang, and Yiming Yang.
\newblock Improve vision language model chain-of-thought reasoning.
\newblock \emph{arXiv preprint arXiv:2410.16198}, 2024.

\bibitem[Zhang et~al.(2022)Zhang, Zhang, Li, and Smola]{zhang2022automatic}
Zhuosheng Zhang, Aston Zhang, Mu~Li, and Alex Smola.
\newblock Automatic chain of thought prompting in large language models.
\newblock \emph{arXiv preprint arXiv:2210.03493}, 2022.

\bibitem[Zhang et~al.(2023)Zhang, Zhang, Li, Zhao, Karypis, and
  Smola]{zhang2023multimodal}
Zhuosheng Zhang, Aston Zhang, Mu~Li, Hai Zhao, George Karypis, and Alex Smola.
\newblock Multimodal chain-of-thought reasoning in language models.
\newblock \emph{arXiv preprint arXiv:2302.00923}, 2023.

\end{thebibliography}
